\documentclass[12pt]{article}
\usepackage{amssymb,amsmath,bm}
\usepackage{epsf}
\usepackage{epsfig,cite}
\usepackage{color}
\def\dfrac#1#2{{\displaystyle {#1 \over #2}}}

\def\simge{\mathrel{\rlap{\raise 0.511ex \hbox{$>$}}{\lower 0.511ex \hbox{$\sim$}}}}
\def\simle{\mathrel{\rlap{\raise 0.511ex \hbox{$<$}}{\lower 0.511ex \hbox{$\sim$}}}}
\def\slash#1{\setbox0=\hbox{$#1$}\dimen0=\wd0
      \setbox1=\hbox{/} \dimen1=\wd1 \ifdim\dimen0>\dimen1
      \rlap{\hbox to \dimen0{\hfil/\hfil}} #1                        \else
      \rlap{\hbox to \dimen1{\hfil$#1$\hfil}}
      /   \fi}

\newcommand{\newsection}[1]{\section{#1}\setcounter{equation}{0}}

\newcommand{\lsim}{
\mathrel{\hbox{\rlap{\hbox{\lower4pt\hbox{$\sim$}}}\hbox{$<$}}}}

\newcommand{\gsim}{
\mathrel{\hbox{\rlap{\hbox{\lower4pt\hbox{$\sim$}}}\hbox{$>$}}}}

\def\eps{\varepsilon}

\newcommand{\tev}{\, {\rm TeV}}
\newcommand{\gev}{\, {\rm GeV}}
\newcommand{\mev}{\, {\rm MeV}}

\allowdisplaybreaks[2]

\newcommand{\be}{\begin{equation}}
\newcommand{\ee}{\end{equation}}
\newcommand{\bea}{\begin{eqnarray}}
\newcommand{\eea}{\end{eqnarray}}

\newcommand{\bi}{\begin{itemize}}
\newcommand{\ei}{\end{itemize}}

\def\kpn{K^+\rightarrow\pi^+\nu\bar\nu}

\def\klpn{K_L\rightarrow\pi^0\nu\bar\nu}

\begin{document}
\begin{titlepage}
\vspace*{-0.5truecm}

%{\Large \today}

\begin{flushright}
TUM-HEP-689/08\\
MPP-2008-50\\
RM3-TH/08-1
\end{flushright}

\vspace{1truecm}

\begin{center}
\boldmath

{\LARGE\textbf{The Littlest Higgs Model with T-Parity \vspace{.2cm}\\
Facing CP-Violation in $B_s - \bar B_s$ Mixing}}

\unboldmath
\end{center}

\vspace{0.4truecm}

\begin{center}
{\bf Monika Blanke$^{a,b}$, Andrzej J.~Buras$^a$, Stefan Recksiegel$^a$, Cecilia  Tarantino$^c$
}
\vspace{0.4truecm}

{\small
 {\sl $^a$Physik Department, Technische Universit\"at M\"unchen,
James-Franck-Stra{\ss}e 2, \\D-85748 Garching, Germany}\vspace{0.2truecm}

 {\sl $^b$Max-Planck-Institut f{\"u}r Physik (Werner-Heisenberg-Institut), 
F\"ohringer Ring 6,\\
D-80805 M{\"u}nchen, Germany}\vspace{0.2truecm}

{\sl $^c$Dipartimento di Fisica, Universit\`a di Roma Tre and INFN, Sez. di Roma Tre,\\
 Via della Vasca Navale 84, I-00146 Roma, Italy}
}

\end{center}

\vspace{0.6cm}
\begin{abstract}
\vspace{0.2cm}\noindent
The non-minimal flavour violating interactions of mirror quarks and new heavy gauge bosons in the Littlest Higgs model with T-parity (LHT) give rise to naturally large CP-violating effects in the $B_s$ system. In view of a large new CP phase in $B_s - \bar B_s$ mixing hinted by the CDF and D{\O} data and the recent UTfit analysis, we update our 2006 analysis of particle-antiparticle mixing and rare $K$ and $B$ decays in the LHT model, using the most recent values of a number of input parameters and performing a more careful error analysis. We find that the CP-asymmetry $S_{\psi\phi}$ can easily reach values $\sim 0.15-0.20$, compared to the SM value $\sim 0.04$, while higher values are rather unlikely though not excluded. Large enhancements are also possible in the branching ratios for $\klpn$, $\kpn$ and $K_L\to\pi^0\ell^+\ell^-$ with much more modest effects in $B_{s,d}\to\mu^+\mu^-$. We perform a detailed study of correlations between the latter decays and  $S_{\psi\phi}$ as well as of the correlation between $S_{\psi\phi}$ and $S_{\psi K_S}$. {We also point out that the possible tension between $\eps_K$ and the tree level CKM determination recently hinted by various analyses can easily be resolved in the LHT model.}

\end{abstract}

\end{titlepage}

\thispagestyle{empty}

\begin{center}
{\Large\bf Note added}
\end{center}

\noindent
An additional contribution to the $Z$ penguin in the Littlest Higgs model with T-parity has been pointed out in \cite{Goto:2008fj,delAguila:2008zu}, which has been overlooked in the present analysis. This contribution leads to the cancellation of the left-over quadratic divergence in the calculation of some rare decay amplitudes. Instead of presenting separate errata to the present work and our papers \cite{Blanke:2006eb,Blanke:2007db,Blanke:2007ee,Blanke:2007wr} partially affected by this omission, we have presented a corrected and updated analysis of flavour changing neutral current processes in the Littlest Higgs model with T-parity in \cite{Blanke:2009am}.

\newpage

\setcounter{page}{1}
\pagenumbering{arabic}

\newsection{Introduction}

One of the most attractive solutions to the so-called {\it little hierarchy
  problem} that affects the Standard Model (SM) is provided by Little Higgs models \cite{Schmaltz:2005ky,Perelstein:2005ka}.
They are perturbatively computable up to $\sim 10 \tev$ and have a rather
  small number of parameters, although their predictivity can be weakened by a
  certain sensitivity to the unknown ultraviolet (UV) completion of the
  theory.
In these models, in contrast to supersymmetry, the problematic quadratic
divergences to the Higgs mass are cancelled by loop contributions of new
particles with the same spin-statistics of the SM ones and with masses around
$1 \tev$.

The basic idea of Little Higgs models\cite{ArkaniHamed:2001nc} is that the Higgs is 
naturally light as it is identified with a Nambu-Goldstone 
boson (NGB) of a spontaneously broken global symmetry.
While an exact NGB would have only derivative interactions, gauge and Yukawa interactions of the Higgs have to be incorporated. This can
be done without generating
quadratically divergent one-loop contributions to the Higgs mass, through the
so-called {\it collective symmetry breaking}. 
The collective symmetry breaking has the peculiarity of generating
the Higgs mass only when two or more couplings in the Lagrangian are
non-vanishing, thus avoiding one-loop quadratic divergences.

The most economical, in matter content, Little Higgs model is the Littlest
Higgs (LH) model\cite{ArkaniHamed:2002qy}, where the global group $SU(5)$ is spontaneously broken
into $SO(5)$ at the scale $f \sim \mathcal{O}(1 \tev)$ and
the electroweak sector of the SM is embedded in an $SU(5)/SO(5)$ non-linear
sigma model. 
Gauge and Yukawa Higgs interactions are introduced by gauging the subgroup of
$SU(5)$: $[SU(2) \times U(1)]_1 \times [SU(2) \times U(1)]_2$. 
In the LH model, the new particles appearing at the $\tev$ scale are the heavy
gauge bosons ($W^\pm_H, Z_H, A_H$), the heavy top ($T$) and the scalar triplet 
$\Phi$.

In the LH model, the
custodial $SU(2)$ symmetry, of fundamental importance for electroweak precision
studies, is unfortunately broken already at tree level, implying
that the relevant scale of New Physics (NP), $f$, must be at least $(2-3)
\tev$ in order to be consistent with electroweak precision data
\cite{Han:2003wu,Csaki:2002qg,Hewett:2002px,Chen:2003fm,Chen:2004ig,Yue:2004xt,Kilian:2003xt}. As a consequence, the original motivation of solving the little hierarchy problem is partly lost. Moreover, the contributions of the new
particles to FCNC processes turn out to be at most $(10-20)\%$
\cite{Buras:2004kq,Choudhury:2004bh,Buras:2005iv,Huo:2003vd,Buras:2006wk}, which will not be easy to distinguish from the
SM due to experimental and theoretical uncertainties.  In particular, 
detailed analyses of particle-antiparticle mixing  and of rare $K$ and $B$ decays in the LH model have been
published in \cite{Buras:2004kq,Buras:2006wk}.

More promising and more interesting from the point of view of FCNC
processes is the Littlest Higgs model with a discrete symmetry called
T-parity~\cite{Cheng:2003ju,Cheng:2004yc,Low:2004xc} under which all new particles listed
above, except $T_+$, are odd and do not contribute to processes
with external SM quarks (T-even) at tree level. As a
consequence, the NP scale $f$ can be lowered down to $1
\tev$ and even below it, without violating electroweak precision
constraints \cite{Hubisz:2005tx,Asano:2006nr}.

A consistent and phenomenologically viable Littlest Higgs model
with T-parity (LHT) requires the introduction of three doublets of
``mirror quarks'' and three doublets of ``mirror leptons'' which
are odd under T-parity, transform vectorially under $SU(2)_L$ and
can be given a large mass. Moreover, there is an additional
heavy $T_-$ quark that is odd under T-parity \cite{Low:2004xc}.

Mirror fermions are characterised by new flavour violating interactions with SM fermions and heavy gauge bosons, which involve in the quark sector two new unitary mixing
matrices, called $V_{Hu}$ and $V_{Hd}$, analogous to the CKM matrix~\cite{Cabibbo:1963yz,Kobayashi:1973fv}.
$V_{Hd}$ contains $3$ angles, like $V_\text{CKM}$, but $3$
  (non-Majorana) phases \cite{Blanke:2006xr}, i.\,e. 
  two additional phases relative to the SM mixing matrices, that cannot be rotated
  away in this case.

Because of these new mixing matrices, the LHT model does not belong to the
Minimal Flavour Violation (MFV) class of models~\cite{Buras:2000dm,Buras:2003jf,D'Ambrosio:2002ex,Chivukula:1987py,Hall:1990ac} and 
significant effects in flavour observables are possible. In particular, the mirror fermion effects on FCNC observables in the quark sector have been studied in detail in \cite{Hubisz:2005bd,Blanke:2006sb,Blanke:2006eb,Blanke:2007ee,Blanke:2007wr,HongSheng:2007ve}, while an extensive analysis of lepton flavour violation has been performed in \cite{Blanke:2007db,Choudhury:2006sq}.

The beauty of the LHT model, when compared with other models with non-minimal 
flavour violating interactions, like the general MSSM and Randall-Sundrum  \cite{Randall:1999ee,Agashe:2004cp} scenarios, is a relatively small
number of new parameters and the fact that the local operators
involved are the same as in the SM. Therefore, the non-perturbative
uncertainties present in certain quantities already in the SM
are the same in the LHT model, and the departures from
the SM are entirely due to short distance physics that can be
calculated within perturbation theory. In stating this we are
aware of the fact that we deal here with an effective field theory
whose ultraviolet completion has not been specified, with the
consequence that at a certain level of accuracy one has to worry
about the effects coming from the cut-off scale $\Lambda\sim 4\pi
f$.

During the last two years several changes in the input parameters involved in our LHT analysis have taken place, and most recently some hints for large CP-violating effects in $b\to s$ transitions  with $\Delta F=2$ have been pointed out. More explicitly:
\begin{enumerate}
\item
The value of $|V_{ub}|$ has been visibly lowered so that presently the tree level determination of the CKM matrix parameters is, in contrast to the situation in 2006, fully compatible with the size of the mixing-induced CP-asymmetry $S_{\psi K_S}$, although a small \emph{negative} phase $\varphi_{B_d}$, defined through
\be
S_{\psi K_S} = \sin(2\beta + 2 \varphi_{B_d}),
\ee
cannot be excluded. Here $\beta$ is the true angle of the unitarity triangle. 
\item
The value of the running top quark mass $\bar m_t(m_t)$ has decreased by roughly $1\gev$.
\item
In the last two years, the lattice calculations of the parameter $\hat B_K$, that governs the CP-violating parameter $\eps_K$, have been performed including for the first time the effects of dynamical quarks \cite{Gamiz:2006sq,Antonio:2007pb,Aoki:2008ss}. An average based on the unquenched results and on the scale dependence suggested by quenched studies reads $\hat B_K = 0.75 \pm 0.07$ \cite{UTfitNEW} that is lower than the central value used in our 2006 analysis. {In addition the analyses in \cite{Antonio:2007pb,Aoki:2008ss} favour values for $\hat B_K$ in the ballpark of 0.70, whose implications on possible NP effects in $\eps_K$ have been discussed in \cite{Lunghi:2008aa,Buras:2008nn}.}
\item
Most interestingly the extracted value of $S_{\psi\phi}$ from the CDF \cite{Aaltonen:2007he} and D{\O} \cite{:2008fj} data turns out to be much larger than the SM value $S_{\psi\phi} \simeq 0.04$. By combining the CDF and D{\O} data, in fact, the UTfit collaboration finds \cite{Bona:2008jn}
\be
\label{eq:Spsiphi}
0.32\le S_{\psi\phi} \le 0.87 \qquad (95\%\text{ C.L.})\,.
\ee
\end{enumerate}

This latter result is certainly of interest for the LHT model as large values of this asymmetry have been found to be allowed by our analysis in \cite{Blanke:2006sb}. Moreover the three effects 1.--3. significantly decrease the value of $\eps_K$ in the SM so that some small contributions from NP in the $K$ system are welcome in order to reproduce the experimental value of $\eps_K$.

All these findings motivate a new analysis of particle-antiparticle mixing and of rare $K$ and $B$ decays. In the present paper we update the most interesting results of our analyses in \cite{Blanke:2006sb,Blanke:2006eb} addressing, in particular, the following questions:
\bi
\item
How large values of $S_{\psi\phi}$ can be obtained in the LHT model consistently with all other available data on FCNC processes?
\item
What would be the impact on the LHT predictions, if any, of a low value for $\hat B_K \simeq 0.70$ as hinted by the recent lattice determinations \cite{Antonio:2007pb,Aoki:2008ss}?
%\item
%What is the correlation, if any, between NP contributions to $S_{\psi\phi}$ and $\eps_K$ in view of the present situation for %these two observables summarised above?
\item
What are the LHT upper bounds on the branching ratios of the rare decays $\klpn$, $\kpn$, $K_L\to\pi^0\ell^+\ell^-$ and $B_s \rightarrow \mu^+ \mu^-$ as functions of $S_{\psi\phi}$?
\item
How strong is in the LHT model the possible violation of the ``golden'' MFV relations between CP-violation in $B_d$-mixing and in rare $K$ decays and between $B_d$ and $B_s$ observables \cite{Buchalla:1994tr,Buras:2003td}, and how does it depend on $S_{\psi\phi}$?
\ei

Our paper is organised as follows. In Section \ref{sec:model} we summarise very briefly the main ingredients of the LHT model, referring frequently to \cite{Blanke:2006sb} and \cite{Blanke:2006eb}, where a more detailed description and all analytical expressions for the observables considered in our numerical analysis can be found. In Section \ref{sec:num}, the main section of the present paper, we answer the questions posed above. We summarise and conclude in Section \ref{sec:conc}.

\newsection{The LHT Model}\label{sec:model}
A detailed description of the LHT model can be found for instance in
\cite{Hubisz:2004ft,Blanke:2006eb}. Here we just want to briefly review the particle content and the flavour structure of the LHT model.

\subsection{Gauge Boson Sector}
\label{subsec:2.1}

The T-even electroweak gauge boson sector \cite{ArkaniHamed:2002qy}  consists only of SM
electroweak gauge bosons
\begin{equation}\label{2.1}
W_L^\pm\,,\qquad Z_L\,,\qquad A_L\,,
\end{equation}
with masses given to lowest order in $v/f$ by
\begin{equation}\label{2.2}
M_{W_L}=\frac{gv}{2}\,,\qquad
M_{Z_L}=\frac{M_{W_L}}{\cos{\theta_W}}\,,\qquad M_{A_L}=0\,,
\end{equation}
where $\theta_W$ is the weak mixing angle. T-parity ensures that
the second relation in \eqref{2.2} is satisfied at tree level to
all orders in $v/f$.

The T-odd gauge boson sector~\cite{ArkaniHamed:2002qy} consists of the three heavy
``partners'' of the SM gauge bosons in \eqref{2.1}:
\begin{equation}\label{2.3}
W_H^\pm\,,\qquad Z_H\,,\qquad A_H\,,
\end{equation}
with masses given to lowest order in $v/f$ by
\begin{equation}\label{2.4}
M_{W_H}=gf\,,\qquad M_{Z_H}=gf\,,\qquad
M_{A_H}=\frac{g'f}{\sqrt{5}}\,,
\end{equation}
that satisfy the relation
\begin{equation}\label{2.4a}
M_{A_H}=\frac{\tan{\theta_W}}{\sqrt{5}}M_{W_H}\simeq\frac{M_{W_H}}{4.1}\,.
\end{equation}

\subsection{Fermion Sector}
\label{subsec:2.2}

The T-even fermion sector~\cite{ArkaniHamed:2002qy}
consists of the SM quarks and leptons and a colour triplet heavy
quark $T_+$ that is, to leading order in $v/f$, singlet under
$SU(2)_L$ and has the mass
\begin{equation}\label{2.5}
m_{T_+}=\frac{f}{v}\frac{m_t}{\sqrt{x_L(1-x_L)}}\,,\quad \quad {\rm with}\quad  x_L =\frac{\lambda_1^2}{\lambda_1^2+\lambda_2^2}\,.
\end{equation}
Here $\lambda_1$ is the Yukawa coupling in the $(t, T_+)$ sector and
$\lambda_2$ parameterises the mass term of $T_+$.

The T-odd fermion sector~\cite{Low:2004xc} consists first of all of three
generations of mirror quarks and leptons with vectorial
couplings under $SU(2)_L$. In this paper only mirror quarks are
relevant. We will denote them by 
  \begin{equation}\label{2.6}
\begin{pmatrix} u^1_{H}\\d^1_{H} \end{pmatrix}\,,\qquad
\begin{pmatrix} u^2_{H}\\d^2_{H} \end{pmatrix}\,,\qquad
\begin{pmatrix} u^3_{H}\\d^3_{H} \end{pmatrix}\,,
\end{equation}
with their masses satisfying to first order in $v/f$
\begin{equation}\label{2.7}
m^u_{H1}=m^d_{H1}\,,\qquad m^u_{H2}=m^d_{H2}\,,\qquad
m^u_{H3}=m^d_{H3}\,.
\end{equation}

The T-odd fermion sector contains also a T-odd heavy quark $T_-$
which, however, does not enter our analysis \cite{Hubisz:2005bd,Blanke:2006sb,Blanke:2006eb}. 

For completeness we mention that also a Higgs
triplet $\Phi$ belongs to the T-odd sector. The charged Higgs
$\phi^\pm$, as well as the neutral Higgses $\phi^0,\;\phi^P$, are
relevant in principle for the decays considered here, but their
effects turn out to be of higher order in $v/f$ \cite{Blanke:2006sb,Blanke:2006eb}, and consequently similarly to $T_-$ will not enter our analysis. 

\subsection{Weak Mixing in the Mirror Sector}
\label{subsec:2.4}
As discussed in detail in~\cite{Hubisz:2005bd,Blanke:2006xr,Blanke:2006sb,Blanke:2006eb}, one of the important
ingredients of the mirror quark sector is the existence of two CKM-like unitary
mixing matrices $V_{Hu}$ and $V_{Hd}$, that satisfy
\begin{equation}\label{2.11}
V_{Hu}^\dagger V_{Hd}=V_\text{CKM}\,.
\end{equation}
These mirror mixing
matrices parameterise flavour violating
interactions between SM fermions and mirror fermions
that are mediated by the heavy gauge bosons $W_H$, $Z_H$ and $A_H$.
The notation indicates the type of light fermion that is involved in the interaction, i.\,e. if it is of up- or down-type.

Following~\cite{Blanke:2006xr} we will parameterise $V_{Hd}$ generalising
the usual CKM parameterisation, as a product of three rotations, and
introducing a complex phase in each of them, thus obtaining
\be\label{2.12a}
\addtolength{\arraycolsep}{3pt}
V_{Hd}= \begin{pmatrix}
c_{12}^d c_{13}^d & s_{12}^d c_{13}^d e^{-i\delta^d_{12}}& s_{13}^d e^{-i\delta^d_{13}}\\
-s_{12}^d c_{23}^d e^{i\delta^d_{12}}-c_{12}^d s_{23}^ds_{13}^d e^{i(\delta^d_{13}-\delta^d_{23})} &
c_{12}^d c_{23}^d-s_{12}^d s_{23}^ds_{13}^d e^{i(\delta^d_{13}-\delta^d_{12}-\delta^d_{23})} &
s_{23}^dc_{13}^d e^{-i\delta^d_{23}}\\
s_{12}^d s_{23}^d e^{i(\delta^d_{12}+\delta^d_{23})}-c_{12}^d c_{23}^ds_{13}^d e^{i\delta^d_{13}} &
-c_{12}^d s_{23}^d e^{i\delta^d_{23}}-s_{12}^d c_{23}^d s_{13}^d
e^{i(\delta^d_{13}-\delta^d_{12})} & c_{23}^d c_{13}^d\\
\end{pmatrix}
\ee
As in the case of the CKM matrix 
the angles $\theta_{ij}^d$ can all be made to lie in the first quadrant 
with $0\le \delta^d_{12}, \delta^d_{23}, \delta^d_{13} < 2\pi$.
The matrix $V_{Hu}$ is then determined through 
$V_{Hu}=V_{Hd}V_\text{CKM}^\dagger$.

\newsection{Numerical Analysis}\label{sec:num}

\subsection{Preliminaries}

%In our numerical analysis we will set $|V_{us}|$, $|V_{cb}|$ and $|V_{ub}|$ to their central values measured in tree level decays~\cite{Barberio:2007cr,Blucher:2005dc} and collected in Table~\ref{tab:input}.

\begin{table}[ht]
\renewcommand{\arraystretch}{1}\setlength{\arraycolsep}{1pt}
\center{\begin{tabular}{|l|l|}
\hline
$\lambda=|V_{us}|= 0.2261(15)$ \hfill \cite{Flavianet}& $G_F= 1.16637\cdot 10^{-5}\gev^{-2}$ \qquad {} \\
$|V_{ub}| = 3.8(4)\cdot 10^{-3}$ \hfill\cite{UTfitNEW} &  $M_W = 80.425 \gev$ \\
$|V_{cb}|= 4.1(1)\cdot 10^{-2}$ \hfill\cite{UTfitNEW}& $\alpha(M_Z) = 1/127.9$ \\\cline{1-1}
$\gamma = 80(20)^\circ $ & $\sin^2\theta_W = 0.23122$\\\cline{1-1}
$\Delta M_K= 0.5292(9)\cdot 10^{-2} \,\text{ps}^{-1}$ \qquad {} & $m_K^0= 497.648\mev$ \\
$|\eps_K|= 2.280(13)\cdot 10^{-3}$ \hfill\cite{Yao:2006px}& $m_{B_d}= 5279.5\mev$ \\\cline{1-1}
$\Delta M_d = 0.507(5) \,\text{ps}^{-1}$ & $m_{B_s} = 5366.4\mev$ \hfill\cite{Yao:2006px} \\\cline{2-2}
$\Delta M_s = 17.77(12) \,\text{ps}^{-1}$  & $\eta_1= 1.32(32)$ \hfill\cite{Herrlich:1993yv}\\\cline{2-2}
$S_{\psi K_S}= 0.681(25)$ \hfill\cite{Barberio:2007cr}&  $\eta_3=0.47(5)$ \hfill\cite{Herrlich:1995hh,Herrlich:1996vf} \\\hline
$\bar m_c = 1.30(5)\gev$ & $\eta_2=0.57(1)$ \\
$\bar m_t = 162.7(13)\gev$ \hfill\cite{Yao:2006px}& $\eta_B=0.55(1)$ \hfill \cite{Buras:1990fn,Urban:1997gw} \\\hline
$f_K = 156(1)\mev$ \hfill \cite{Flavianet} & $f_{B_s} = 245(25)\mev$\\\cline{1-1}
$\hat B_K= 0.75(7)$ & $f_{B_d} = 200(20)\mev$ \\
$\hat B_{B_s} = 1.22(12)$ & $f_{B_s} \sqrt{\hat B_{B_s}} = 270(30)\mev$ \\
$\hat B_{B_d} = 1.22(12)$ & $f_{B_d} \sqrt{\hat B_{B_d}} = 225(25)\mev$ \\
$\hat B_{B_s}/\hat B_{B_d} = 1.00(3)$ \hfill \cite{UTfitNEW} & $\xi = 1.21(4)$ \hfill \cite{UTfitNEW}
 \\\hline
\end{tabular}  }
\caption {\textit{Values of the experimental and theoretical
    quantities used as input parameters.} }
\label{tab:input}
\renewcommand{\arraystretch}{1.0}
\end{table}

In our previous analyses of flavour physics observables \cite{Blanke:2006sb,Blanke:2006eb,Blanke:2007ee,Blanke:2007wr}, we simplified the numerical analysis by setting all input parameters to their central values and allowing instead $\Delta M_K$, $\varepsilon_K$, $\Delta M_d$, $\Delta M_s$, $\Delta M_s/\Delta M_d$ and $S_{\psi K_S}$ to differ from 
their experimental values by $\pm 50\%$, $\pm 40\%$, $\pm 40\%$, $\pm 40\%$,
$\pm 20\%$ and $\pm 8\%$, respectively. While this simplifying assumption was
justified in order to determine the size of possible NP
effects in observables that have not been observed so far, an improved error analysis
is now required in order to be able to draw more accurate remarks, in view of
the recent significant improvements both of the experimental constraints on the
NP phase $\varphi_{B_s}$ in the $B_s$ system {\cite{Aaltonen:2007he,:2008fj,Bona:2008jn}} and of
the lattice determinations of the non-perturbative parameter $\hat B_K$
  \cite{Gamiz:2006sq,Antonio:2007pb,Aoki:2008ss}. Therefore, in
what follows, we will take all input parameters to be flatly distributed within
their $1\sigma$ ranges indicated in Table \ref{tab:input}. 
At the same time we require the observables              
$\varepsilon_K$, $\Delta M_d$, $\Delta M_s$ and $S_{\psi K_S}$, resulting
from SM and LHT contributions, to lie within their experimental $1\sigma$
ranges. In the case of $\Delta M_K$ where the theoretical uncertainty is
large due to unknown long-distance contributions, we allow the generated  
value to lie within $\pm 30\%$ of the experimental central value.
%{\bf Although this strategy cannot compete with a global $\chi^2$ fit to all observables, we believe that such an analysis will be necessary only when more observables have been measured and the errors in the input parameters are significantly decreased.}
%
All formulae for the observables discussed in the present paper can be found in \cite{Blanke:2006sb,Blanke:2006eb}. 

In \cite{Blanke:2006sb,Blanke:2006eb} several benchmark scenarios for the
structure of the $V_{Hd}$ matrix have been discussed. Here we confine our
discussion only to the general scan, where we perform, as in
\cite{Blanke:2006eb}, a scan over the mirror fermion masses and the $V_{Hd}$
parameters, with the NP scale $f$ fixed to 1 TeV. 
To achieve this, we generate a large number of points where all
mirror fermion masses are varied in the interval [300GeV,1TeV],
all angles in the interval $[0,\pi/2]$, all phases between 0 and 2$\pi$ and all SM input 
parameters are varied in their $1\sigma$ ranges. Our plots 
show a sample of 15000 points that fulfil
all experimental constraints. We do not specifically filter for
``interesting'' points, therefore the point density gives us an
idea of how likely it is for the LHT model to generate a certain effect.
To have a global view of the most general LHT effects, we have allowed
  here the phases $\delta_{12}^d$ and $\delta_{23}^d$ to differ from zero.
Qualitatively their effect is not significant, although they can help in
achieving very large effects in certain observables. 
In \cite{Blanke:2006eb} we found that there exist some sets of masses and $V_{Hd}$
parameters where the NP effects turn out
to be spectacular in both $B$ and $K$ systems.
As we will see below, the present analysis, where a more accurate treatment of the uncertainties has been performed with respect to \cite{Blanke:2006sb,Blanke:2006eb}, still permits large departures from the SM expectations.

\subsection{Results}
In all the plots presented here, we show only points consistent {with the experimental $1\sigma$ ranges of $\Delta F = 2$ data, in particular with $\eps_K$ and $S_{\psi K_S}$. We}
% of $S_{\psi K_S}$ and we
represent the SM predictions and the LHT T-even contributions by black and light-blue dots, respectively.

%%%%%%%%%%%%%%%%%%%%%%%%%%%%%%%%%%%%%%%%%%%%%%%%%%%%%%%%%%%%%%
\begin{figure}
\center{\epsfig{file=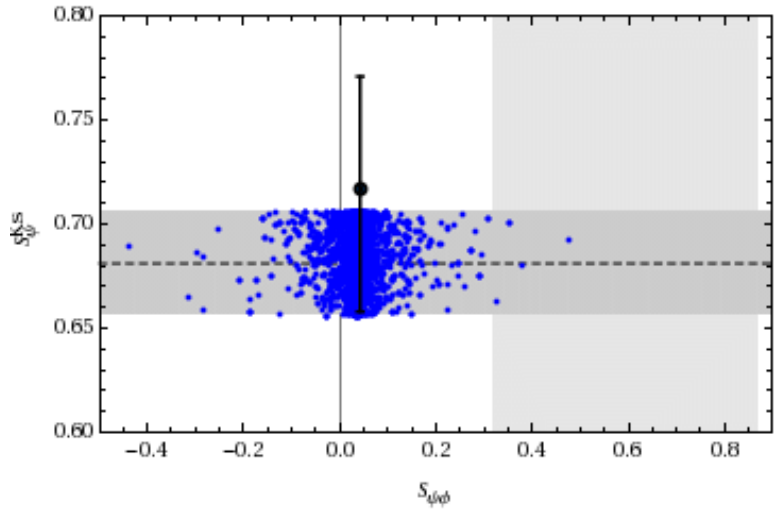,scale=.8}}
\caption{\textit{$S_{\psi K_S}$ as a function of $S_{\psi\phi}$. The black dot represents the tree-level SM prediction (see text), whose uncertainty is dominated by the error on $V_{ub}$, {shown as error bar. The grey horizontal band displays the experimental $1\sigma$ range for $S_{\psi K_S}$, while the range given in \eqref{eq:Spsiphi} is shown by the light-grey vertical band.}}}
\label{fig:SpsiphiSpsiKS}
\end{figure}
%%%%%%%%%%%%%%%%%%%%%%%%%%%%%%%%%%%%%%%%%%%%%%%%%%%%%%%%%%%%%%
In Fig.~\ref{fig:SpsiphiSpsiKS} we show the CP-asymmetry $S_{\psi K_S}$ versus $S_{\psi\phi}$. 
We observe that the SM prediction for $S_{\psi K_S}$ (black dot), obtained from the tree-level unitarity triangle analysis with the input parameters given by the central values in Table \ref{tab:input}, is around $1.5\sigma$ higher than the data.
On the other hand, when one considers the uncertainties on the input parameters, the SM prediction for $S_{\psi K_S}$ is found to be in good agreement with the data, showing that
the tension between the data on $V_{ub}$ and $S_{\psi K_S}$, previously known as the ``$\sin 2 \beta$ problem" \cite{Bona:2007mn,Buras:2005hz,Blanke:2006ig,Blanke:2006sb}, has disappeared.
{In order to illustrate this, we show for the SM prediction for $S_{\psi K_S}$ also the error bar originating from the uncertainty in $V_{ub}$, which turns out to be the dominant one.
}
In the LHT model we find that sizable both positive and negative values of $S_{\psi\phi}$ relative to the SM value $(S_{\psi\phi})_{\text{SM}}\simeq 0.04$ are possible, and values as high as $\sim 0.15-0.20$ can easily be reached.
While higher values are rather unlikely, they are not excluded at present.
In addition a reversal of the sign of $S_{\psi\phi}$ appears to be unlikely, in accordance with recent data.
On the other hand, basically all values for $S_{\psi\phi}$ found in the LHT model are outside the range \eqref{eq:Spsiphi}, and consequently the confirmation of this result would put the LHT model into difficulties.

%%%%%%%%%%%%%%%%%%%%%%%%%%%%%%%%%%%%%%%%%%%%%%%%%%%%%%%%%%%%%%
\begin{figure}
\center{\epsfig{file=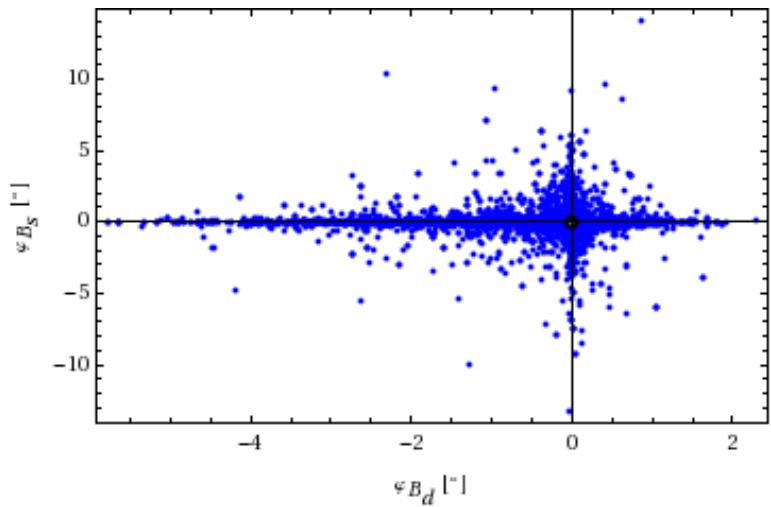,scale=.8}}
\caption{\textit{The $B^0_s-\bar B^0_s$ phase $\varphi_{B_s}$ as a function of the $B^0_d-\bar B^0_d$ phase $\varphi_{B_d}$.}}
\label{fig:phiBdphiBs}
\end{figure}
%%%%%%%%%%%%%%%%%%%%%%%%%%%%%%%%%%%%%%%%%%%%%%%%%%%%%%%%%%%%%%
In Fig.~\ref{fig:phiBdphiBs} we show the allowed points in the $(\varphi_{B_d},\varphi_{B_s})$ plane.
We note, again, that $\varphi_{B_d}<0$ is preferred to fit the $S_{\psi K_S}$ data, while $\varphi_{B_s}$ appears rather symmetric around zero, with $\varphi_{B_s}<0$ favoured by the recent CDF and D{\O} data \cite{Aaltonen:2007he,:2008fj,Bona:2008jn}, being \cite{Blanke:2006ig}
\be
S_{\psi\phi} = \sin(2|\beta_s| -2\varphi_{B_s})\,.
\ee 
We observe, in particular, that in the LHT model $|\varphi_{B_s}| < 10^\circ$ which is significantly lower than $|\varphi_{B_s}| \simeq 20^\circ$ corresponding to the {central value of \eqref{eq:Spsiphi}.
%We also find that large values of $\varphi_{B_s}$ can be obtained independently of the size of the LHT contribution to $\eps_K$.
Compared to} our results in \cite{Blanke:2006sb,Blanke:2006eb}, this improved error analysis and the modified input parameters do not significantly lower the maximal NP effects in $S_{\psi\phi}$, but make values $S_{\psi\phi}\gsim 0.2$ more unlikely.

{
In addition, we analysed possible correlations between the NP phase $\varphi_{B_s}$ and the LHT contributions to $\eps_K$, where we found no relevant correlation. Therefore, the confirmation of a low value of $\hat B_K\simeq0.70$ hinted by recent lattice determinations \cite{Antonio:2007pb,Aoki:2008ss} would not significantly modify our present conclusions. Indeed, as the contributions from the T-even sector always enhance $\eps_K$ with respect to its SM value \cite{Blanke:2006sb}, the LHT model would be welcome to cure a possible tension between $\eps_K$ and the tree level determined $\sin2\beta_\text{true}$, recently hinted in \cite{Lunghi:2008aa,Buras:2008nn}.}

%%%%%%%%%%%%%%%%%%%%%%%%%%%%%%%%%%%%%%%%%%%%%%%%%%%%%%%%%%%%%%
\begin{figure}
\center{\epsfig{file=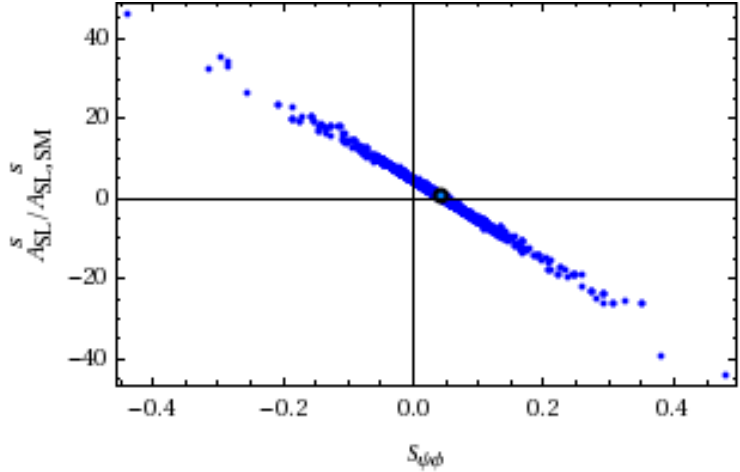,scale=.9}}
\caption{\textit{The semileptonic CP-asymmetry $A^s_\text{SL}$ normalised to its SM central value as a function of $S_{\psi\phi}$.}}
\label{fig:AsSL}
\end{figure}
%%%%%%%%%%%%%%%%%%%%%%%%%%%%%%%%%%%%%%%%%%%%%%%%%%%%%%%%%%%%%%
In Fig.~\ref{fig:AsSL} we show the correlation between $A^s_\text{SL}$ normalised to its SM central value versus $S_{\psi\phi}$. This plot is similar to the one in \cite{Blanke:2006sb}, but again huge enhancements of $A^s_\text{SL}$ and $S_{\psi\phi}$ are less likely than found in our 2006 analysis.

%%%%%%%%%%%%%%%%%%%%%%%%%%%%%%%%%%%%%%%%%%%%%%%%%%%%%%%%%%%%%%
\begin{figure}
\center{\epsfig{file=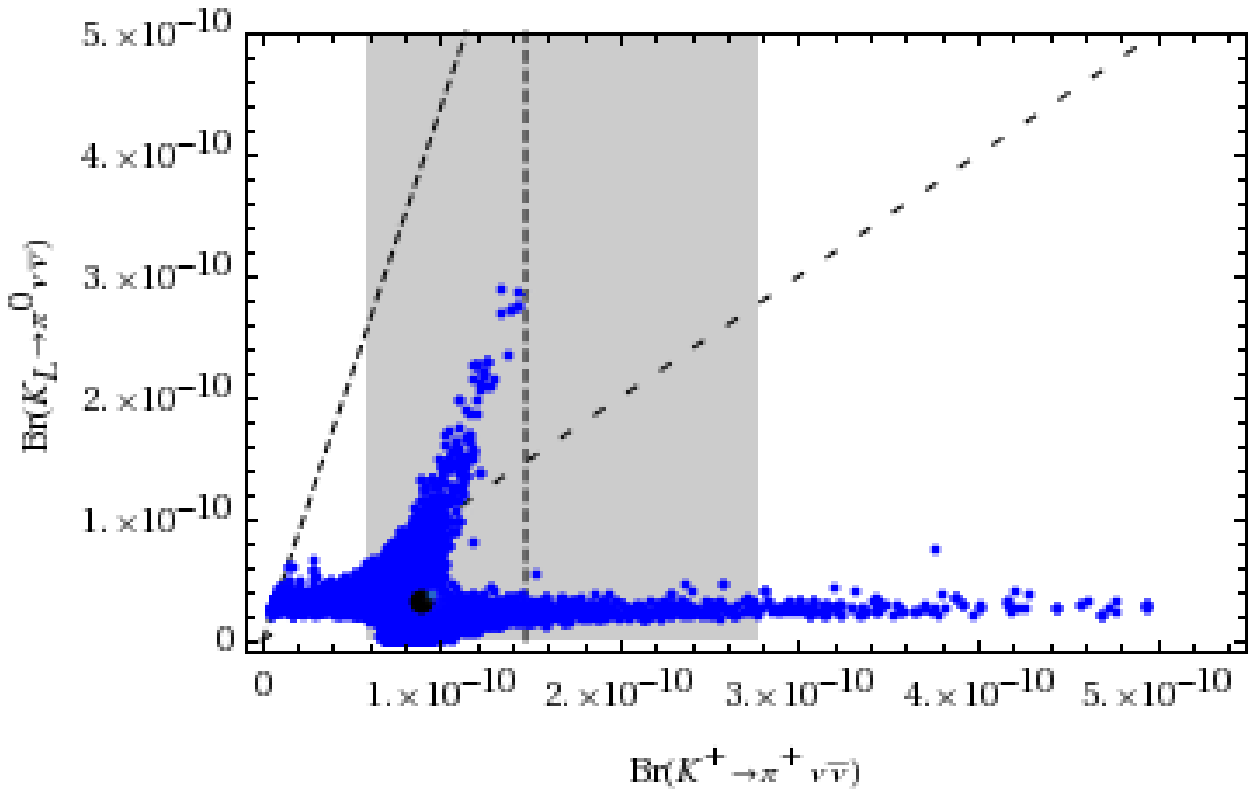,scale=.8}}
\caption{\textit{The branching ratio $Br(\klpn)$ as a function of $Br(\kpn)$.}}
\label{fig:Kpnn}
\end{figure}
%%%%%%%%%%%%%%%%%%%%%%%%%%%%%%%%%%%%%%%%%%%%%%%%%%%%%%%%%%%%%%
In Fig.~\ref{fig:Kpnn} we show the correlation between $Br(\kpn)$ and $Br(\klpn)$. The {experimental $1\sigma$ range} for $Br(\kpn)$ \cite{Adler:2001xv,Anisimovsky:2004hr} and the model-independent Grossman-Nir (GN) bound \cite{Grossman:1997sk} are also shown. We observe that the two branches of possible points identified in \cite{Blanke:2006eb} still appear. The first one is parallel to the GN-bound and leads to possible large enhancements of $Br(\klpn)$ up to values as high as $3\cdot 10^{-10}$, being perfectly consistent with the measured value for $Br(\kpn)$. The second branch corresponds to $Br(\klpn)$ being rather close to its SM prediction, while $Br(\kpn)$ is allowed to vary in the range $[1\cdot 10^{-11},5\cdot 10^{-10}]$, however values above $4\cdot 10^{-10}$ are experimentally disfavoured. We also note, in accordance with our previous findings and in contrast to the SM relation $Br(\klpn)\simeq Br(\kpn)/3$, that  in the LHT model $Br(\klpn)$ can exceed $Br(\kpn)$.
In order to determine how these enhancements would be reduced in the case of a higher NP scale, we have performed an analysis with $f=3 \tev$, finding that $Br(\klpn)$ can be at most enhanced up to {$7\cdot 10^{-11}$, i.\,e. roughly} a factor 5 less relative to the $f= 1 \tev$ case.

%%%%%%%%%%%%%%%%%%%%%%%%%%%%%%%%%%%%%%%%%%%%%%%%%%%%%%%%%%%%%%
\begin{figure}[t!]
\center{\epsfig{file=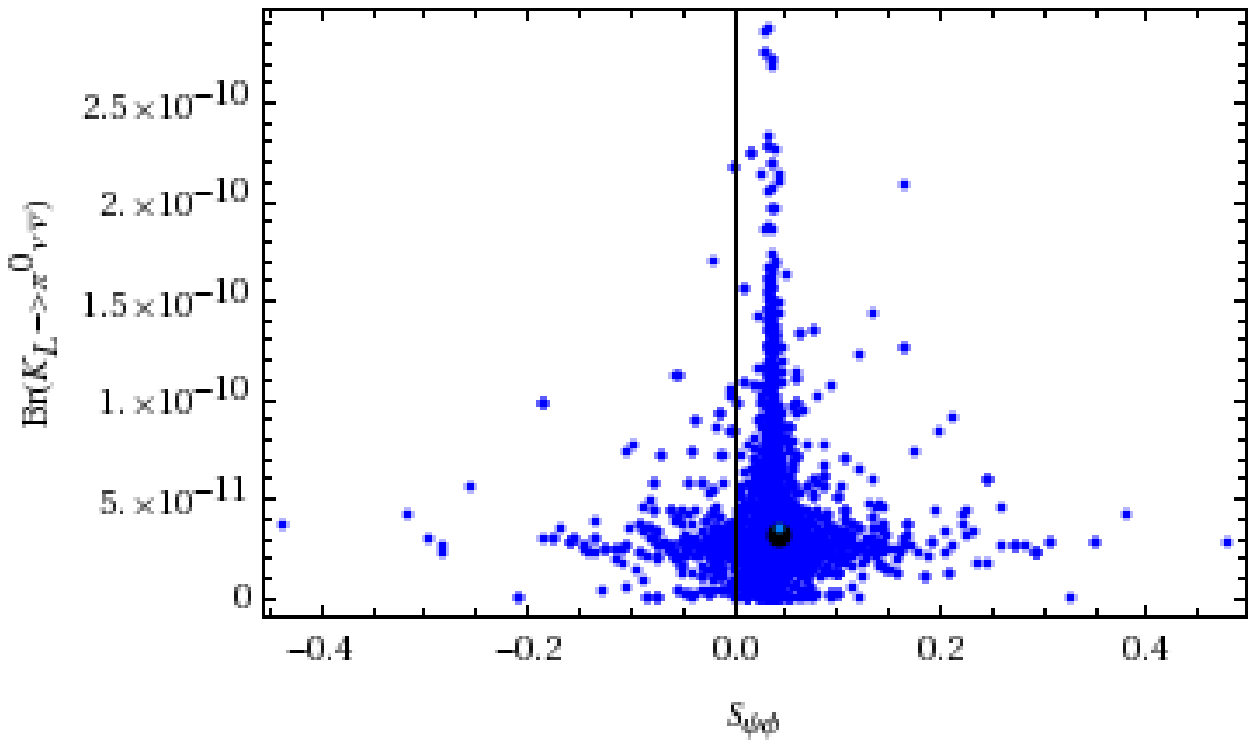,scale=.8}}
\caption{\textit{The branching ratio $Br(\klpn)$ as a function of $S_{\psi\phi}$.}}
\label{fig:KLSpsiphi}
\end{figure}
%%%%%%%%%%%%%%%%%%%%%%%%%%%%%%%%%%%%%%%%%%%%%%%%%%%%%%%%%%%%%%
%%%%%%%%%%%%%%%%%%%%%%%%%%%%%%%%%%%%%%%%%%%%%%%%%%%%%%%%%%%%%%
\begin{figure}[t!]
\center{\epsfig{file=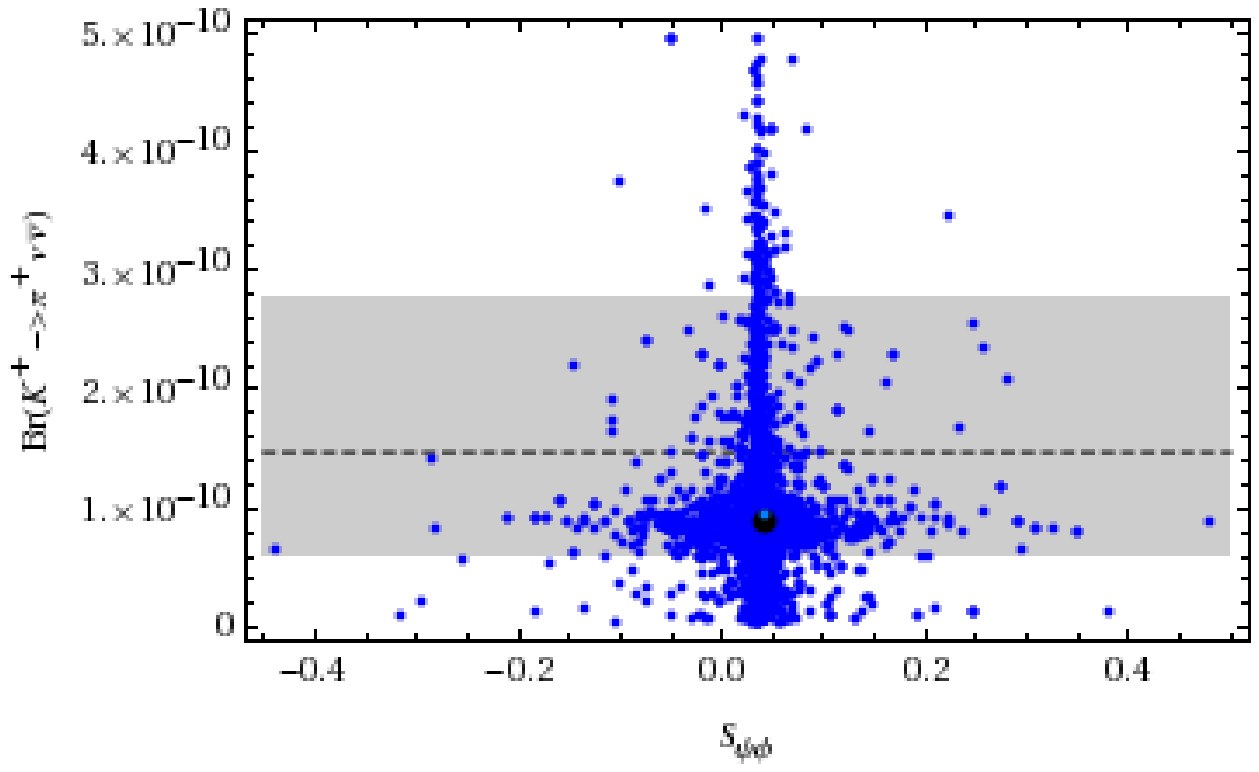,scale=.8}}
\caption{\textit{The branching ratio $Br(\kpn)$ as a function of $S_{\psi\phi}$.}}
\label{fig:K+Spsiphi}
\end{figure}
%%%%%%%%%%%%%%%%%%%%%%%%%%%%%%%%%%%%%%%%%%%%%%%%%%%%%%%%%%%%%%
In Figs.~\ref{fig:KLSpsiphi} and \ref{fig:K+Spsiphi} we show the correlation between $S_{\psi\phi}$ and $Br(\klpn)$ and $Br(\kpn)$, respectively. We observe that large simultaneous enhancements of $S_{\psi\phi}$ and $Br(\klpn)$ are rather unlikely, although for $S_{\psi\phi}\simeq 0.2$ a factor 3 enhancement of $Br(\klpn)$ with respect to its SM value $3\cdot 10^{-11}$ is possible. For higher values of $S_{\psi\phi}$, $Br(\klpn)$ is expected to be SM-like in the LHT model. Consequently, a precise measurement of $S_{\psi\phi}$ will have an important impact on the allowed range for $Br(\klpn)$.
Similarly, {for $S_{\psi\phi}\gsim 0.2$}, the largest enhancements of $Br(\kpn)$ are not allowed, but values as high as {$3\cdot 10^{-10}$} are still possible.

Our new results for the correlation of $K_L\to\pi^0\mu^+\mu^-$ with $K_L\to\pi^0e^+e^-$ are very similar to those found in \cite{Blanke:2006eb}, therefore, here we just mention that $Br(K_L\to\pi^0\mu^+\mu^-)$ and $Br(K_L\to\pi^0e^+e^-)$ can be enhanced up to $3\cdot 10^{-11}$ and {$7\cdot 10^{-11}$}, respectively.

Then, we consider two theoretically clean ratios that are equal to unity in the SM and in MFV models, the so-called ``golden" MFV relations \cite{Buchalla:1994tr,Buras:2001af,Buras:2003td}. Their deviation from one, therefore, would signal an evident effect of NP beyond MFV.
As we discuss below, significant deviations are allowed in the LHT model and could be measured in the near future.

\begin{figure}
\center{\epsfig{file=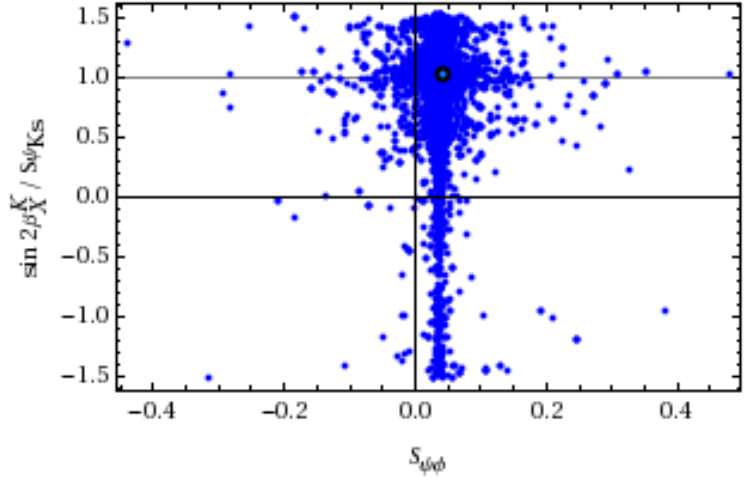,scale=.9}}
\caption{\textit{The ratio $\sin 2 \beta_X^K/S_{\psi K_S}$ as a function of $S_{\psi\phi}$.}}
\label{fig:sin2b}
\end{figure}

In Fig.~\ref{fig:sin2b} we show, as a function of $S_{\psi\phi}$, the ratio $\sin 2 \beta_X^K/S_{\psi K_S}$, where $\beta_X^K$ denotes the angle $\beta$ determined from the $K \to \pi \nu \bar \nu$ system.
In the SM and in MFV models $S_{\psi K_S}$ and $\sin 2 \beta_X^K$ both provide a direct measurement of $\sin 2 \beta$ and their ratio is then equal to one.
Beyond MFV, instead, NP phases can affect the two determinations in a different way.
As we can see from Fig.~\ref{fig:sin2b}, in the LHT model, an enhancement of $50$\% and an even stronger suppression are allowed.
Moreover, a sign inversion is less likely but not excluded. 
For $S_{\psi\phi}\gsim0.1$, very strong suppressions are unlikely, while $\sim 50$\% effects are still possible.

\begin{figure}
\center{\epsfig{file=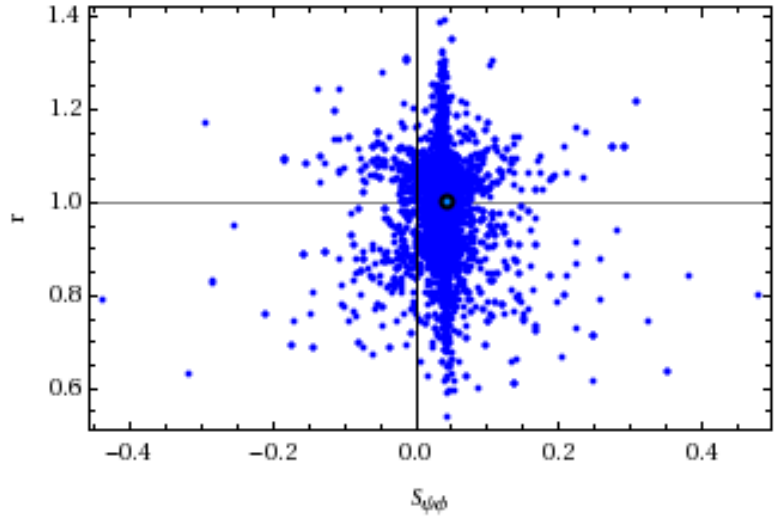,scale=.9}}
\caption{\textit{The ratio $r$, defined in~\eqref{eq:r}, as a function of $S_{\psi\phi}$.}}
\label{fig:r}
\end{figure}

Another interesting parameter that would clearly reveal the presence of NP beyond {constrained MFV (CMFV) \cite{Buras:2000dm,Buras:2003jf,Blanke:2006ig}} is the ratio $r$ defined by \cite{Buras:2003td}
\be
\dfrac{Br(B_s \to \mu^+ \mu^-)}{Br(B_d \to \mu^+ \mu^-)}=\dfrac{\hat B_{B_d}}{\hat B_{B_s}}\dfrac{\tau (B_s)}{\tau (B_d)}\dfrac{\Delta M_s}{\Delta M_d}\,r\,.
\label{eq:r}
\ee
This correlation relates the ratios $Br(B_s \to \mu^+ \mu^-)/Br(B_d \to \mu^+ \mu^-)$ and $\Delta M_s/\Delta M_d$.
In the SM and in {CMFV} models it is valid with $r=1$, while a value of $r$ different from one would signal a NP effect beyond {CMFV}.
As we can see from Fig.~\ref{fig:r}, in the LHT model, values in the range $[0.6,1.3]$ are possible and are more probable if  $S_{\psi\phi}$ is close to the SM value.
%For $S_{\psi\phi}\gsim0.15$, instead, the MFV violation turns out to be smaller, $\lsim 20$\%.
%%%%%%%%%%%%%%%%%%%%%%%%%%%%%%%%%%%%%%%%%%%%%%%%%%%%%%%%%%%%%%
%%%%%%%%%%%%%%%%%%%%%%%%%%%%%%%%%%%%%%%%%%%%%%%%%%%%%%%%%%%%%%
%%%%%%%%%%%%%%%%%%%%%%%%%%%%%%%%%%%%%%%%%%%%%%%%%%%%%%%%%%%%%%
%%%%%%%%%%%%%%%%%%%%%%%%%%%%%%%%%%%%%%%%%%%%%%%%%%%%%%%%%%%%%%

\begin{figure}
\center{\epsfig{file=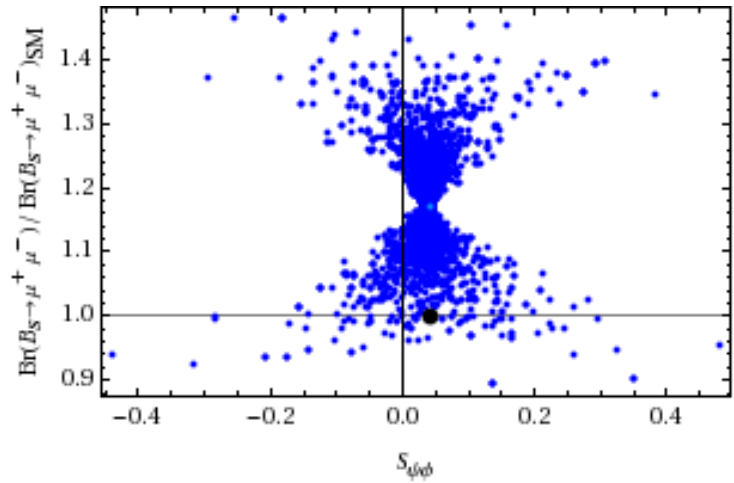,scale=.9}}
\caption{\textit{The ratio $Br(B_s \to \mu^+ \mu^-)/Br(B_s \to \mu^+ \mu^-)_{\text{SM}}$ as a function of $S_{\psi\phi}$.}}
\label{fig:Bsmu+mu-}
\end{figure}

Finally, we show in Fig.~\ref{fig:Bsmu+mu-} the ratio $Br(B_s \to \mu^+ \mu^-)/Br(B_s \to \mu^+ \mu^-)_{\text{SM}}$ as a function of $S_{\psi\phi}$.
A deviation from one would represent a NP signal that could be measured in future experiments.
As we can see from Fig.~\ref{fig:Bsmu+mu-}, in the LHT model the branching ratio $Br(B_s \to \mu^+ \mu^-)$ tends to be larger than the SM value, mainly due to the T-even contribution (denoted by the light-blue point in the plot).
Large ($\gsim 30$\%) enhancements, however, are found to be very unlikely. 
%%%%%%%%%%%%%%%%%%%%%%%%%%%%%%%%%%%%%%%%%%%%%%%%%%%%%%%%%%%%%%
%%%%%%%%%%%%%%%%%%%%%%%%%%%%%%%%%%%%%%%%%%%%%%%%%%%%%%%%%%%%%%
{In addition it is interesting to note that large NP effects in $S_{\psi\phi}$ coincide with non-vanishing T-odd contributions to $Br(B_s\to\mu^+\mu^-)$, leading to either an additional enhancement or a suppression compensating the T-even effect $Br(B_s\to\mu^+\mu^-)_\text{T-even}\simeq 1.2 Br(B_s\to\mu^+\mu^-)_\text{SM}$, where
\be
Br(B_s\to\mu^+\mu^-)_\text{SM} = (3.61 \pm 0.39)\cdot 10^{-9}\,,
\ee
that we update here for completeness. In total, in the LHT model values as high as  $Br(B_s\to\mu^+\mu^-)\simeq 5.5 \cdot 10^{-9}$ can be reached.} 

\newsection{Conclusions}\label{sec:conc}

In the present paper we have reanalysed {in the LHT model} the most interesting FCNC observables in the quark sector, paying particular attention to the possible implications of a large value of $S_{\psi\phi}$, as hinted by the analysis of \cite{Bona:2008jn}. Our main findings are as follows:
\begin{enumerate}
\item
The CP-asymmetry $S_{\psi\phi}$ in the LHT model can reach values up to
$S_{\psi\phi} \simeq 0.4$, although values above 0.2 appear to be unlikely.
In addition we find no significant correlation of $S_{\psi\phi}$ with the NP
contribution to $\eps_K$. {In particular a possible tension between $\eps_K$ and the tree level $\sin2\beta_\text{true}$ can easily be accommodated in the LHT model.}
\item
For values $S_{\psi\phi}\gsim0.2$ large enhancements of the rare decay branching ratios $Br(\kpn)$, $Br(\klpn)$ and $Br(K_L\to\pi^0\ell^+\ell^-)$ are rather improbable. However, our general scan shows that with $S_{\psi\psi}\simeq 0.2$, $Br(\kpn)$, $Br(\klpn)$ and $Br(K_L\to\pi^0\ell^+\ell^-)$ can still be enhanced {by factors of 3, 3 and 1.5}, respectively.
\item
We also find a strong correlation between $A^s_\text{SL}$ and $S_{\psi\phi}$, such that $A^s_\text{SL}$ can largely deviate from the SM prediction.
%To a good approximation
%\be
%\frac{A^s_\text{SL}}{(A^s_\text{SL})^\text{SM}} = \kappa S_{\psi\phi}\,.
%\ee
\item
The MFV ``golden" relations $\sin 2 \beta_X^K/S_{\psi K_S}=1$ and $r=1$ {in \eqref{eq:r}} can be significantly violated. Deviations of $50$\% and $30$\%, respectively, turn out to be likely in the LHT model.
\item
The branching ratio $Br(B_s \to \mu^+ \mu^-)$ can be enhanced in the LHT model {by at most} {$40$\%} relative to the SM. A $Br(B_s \to \mu^+ \mu^-)$ measurement above {$6 \cdot 10^{-9}$}, therefore, would put the LHT model in difficulties. 
\end{enumerate}

\subsection*{Acknowledgements}
{We would like to thank Bj\"orn Duling and Paride Paradisi for useful discussions.} 
This research was partially supported by {the Cluster of Excellence `Origin and Structure of the Universe' and by} the German `Bundesministerium f{\"u}r Bildung und
Forschung' under contract 05HT6WOA.

\subsection*{Note}
A version of this paper with full resolution plots is available at\\
{\tt http://users.physik.tu-muenchen.de/mblanke/LHT-2008.pdf}.

\providecommand{\href}[2]{#2}\begingroup\raggedright\endgroup

%\bibliography{references}
%\bibliography{/home/cecilia/tmp/LHT/Update/references}
%\bibliography{/home/mblanke/Forschung/references}

\providecommand{\href}[2]{#2}\begingroup\raggedright\begin{thebibliography}{10}




%\cite{Goto:2008fj}
\bibitem{Goto:2008fj}
  T.~Goto, Y.~Okada and Y.~Yamamoto,
  {\it Ultraviolet divergences of flavor changing amplitudes in the littlest Higgs
  model with T-parity},
  Phys.\ Lett.\  B {\bf 670} (2009) 378
  [arXiv:0809.4753 [hep-ph]].
  %%CITATION = PHLTA,B670,378;%%

%\cite{delAguila:2008zu}
\bibitem{delAguila:2008zu}
  F.~del Aguila, J.~I.~Illana and M.~D.~Jenkins,
{\it Precise limits from lepton flavour violating processes on the Littlest
  Higgs model with T-parity},
  JHEP {\bf 0901} (2009) 080
  [arXiv:0811.2891 [hep-ph]].
  %%CITATION = JHEPA,0901,080;%%


%\cite{Blanke:2006eb}
\bibitem{Blanke:2006eb}
  M.~Blanke, A.~J.~Buras, A.~Poschenrieder, S.~Recksiegel, C.~Tarantino, S.~Uhlig and A.~Weiler,
  {\it Rare and CP-Violating $K$ and $B$ Decays in the Littlest Higgs Model with
  T-Parity},
  JHEP {\bf 0701}, 066 (2007)
  [arXiv:hep-ph/0610298].
  %%CITATION = JHEPA,0701,066;%%

%\cite{Blanke:2007db}
\bibitem{Blanke:2007db}
  M.~Blanke, A.~J.~Buras, B.~Duling, A.~Poschenrieder and C.~Tarantino,
 {\it Charged Lepton Flavour Violation and $(g-2)_\mu$ in the Littlest Higgs
  Model with T-Parity: A Clear Distinction from Supersymmetry},
  JHEP {\bf 0705}, 013 (2007)
  [arXiv:hep-ph/0702136].
  %%CITATION = JHEPA,0705,013;%%

%\cite{Blanke:2007ee}
\bibitem{Blanke:2007ee}
  M.~Blanke, A.~J.~Buras, S.~Recksiegel, C.~Tarantino and S.~Uhlig,
  {\it Littlest Higgs Model with T-Parity Confronting the New Data on $D^0 -
  \bar{D}^0$ Mixing},
  Phys.\ Lett.\  B {\bf 657}, 81 (2007)
  [arXiv:hep-ph/0703254].
  %%CITATION = PHLTA,B657,81;%%

%\cite{Blanke:2007wr}
\bibitem{Blanke:2007wr}
  M.~Blanke, A.~J.~Buras, S.~Recksiegel, C.~Tarantino and S.~Uhlig,
  {\it Correlations between $\varepsilon'/\varepsilon$ and Rare $K$ Decays in the Littlest Higgs Model with T-Parity},
  JHEP {\bf 0706}, 082 (2007)
  [arXiv:0704.3329 [hep-ph]].
  %%CITATION = JHEPA,0706,082;%%



%\cite{Blanke:2009am}
\bibitem{Blanke:2009am}
  M.~Blanke, A.~J.~Buras, B.~Duling, S.~Recksiegel and C.~Tarantino,
  {\it FCNC Processes in the Littlest Higgs Model with T-Parity: an Update},
  arXiv:0906.5454 [hep-ph].
  %%CITATION = ARXIV:0906.5454;%%


\bibitem{Schmaltz:2005ky}
M.~Schmaltz and D.~Tucker-Smith, {\it {Little Higgs review}},  {\em Ann. Rev.
  Nucl. Part. Sci.} {\bf 55} (2005) 229--270,
  [\href{http://xxx.lanl.gov/abs/hep-ph/0502182}{{\tt hep-ph/0502182}}].

\bibitem{Perelstein:2005ka}
M.~Perelstein, {\it {Little Higgs models and their phenomenology}},  {\em Prog.
  Part. Nucl. Phys.} {\bf 58} (2007) 247--291,
  [\href{http://xxx.lanl.gov/abs/hep-ph/0512128}{{\tt hep-ph/0512128}}].

\bibitem{ArkaniHamed:2001nc}
N.~Arkani-Hamed, A.~G. Cohen, and H.~Georgi, {\it {Electroweak symmetry
  breaking from dimensional deconstruction}},  {\em Phys. Lett.} {\bf B513}
  (2001) 232--240, [\href{http://xxx.lanl.gov/abs/hep-ph/0105239}{{\tt
  hep-ph/0105239}}].

\bibitem{ArkaniHamed:2002qy}
N.~Arkani-Hamed, A.~G. Cohen, E.~Katz, and A.~E. Nelson, {\it {The littlest
  Higgs}},  {\em JHEP} {\bf 07} (2002) 034,
  [\href{http://xxx.lanl.gov/abs/hep-ph/0206021}{{\tt hep-ph/0206021}}].

\bibitem{Han:2003wu}
T.~Han, H.~E. Logan, B.~McElrath, and L.-T. Wang, {\it {Phenomenology of the
  little Higgs model}},  {\em Phys. Rev.} {\bf D67} (2003) 095004,
  [\href{http://xxx.lanl.gov/abs/hep-ph/0301040}{{\tt hep-ph/0301040}}].

\bibitem{Csaki:2002qg}
C.~Csaki, J.~Hubisz, G.~D. Kribs, P.~Meade, and J.~Terning, {\it {Big
  corrections from a little Higgs}},  {\em Phys. Rev.} {\bf D67} (2003) 115002,
  [\href{http://xxx.lanl.gov/abs/hep-ph/0211124}{{\tt hep-ph/0211124}}].

\bibitem{Hewett:2002px}
J.~L. Hewett, F.~J. Petriello, and T.~G. Rizzo, {\it {Constraining the littlest
  Higgs}},  {\em JHEP} {\bf 10} (2003) 062,
  [\href{http://xxx.lanl.gov/abs/hep-ph/0211218}{{\tt hep-ph/0211218}}].

\bibitem{Chen:2003fm}
M.-C. Chen and S.~Dawson, {\it {One-loop radiative corrections to the rho
  parameter in the littlest Higgs model}},  {\em Phys. Rev.} {\bf D70} (2004)
  015003, [\href{http://xxx.lanl.gov/abs/hep-ph/0311032}{{\tt
  hep-ph/0311032}}].

\bibitem{Chen:2004ig}
M.-C. Chen and S.~Dawson, {\it {The littlest Higgs model and one-loop
  electroweak precision constraints}},
  \href{http://xxx.lanl.gov/abs/hep-ph/0409163}{{\tt hep-ph/0409163}}.

\bibitem{Yue:2004xt}
C.-x. Yue and W.~Wang, {\it {The branching ratio $R_b$ in the littlest Higgs
  model}},  {\em Nucl. Phys.} {\bf B683} (2004) 48--66,
  [\href{http://xxx.lanl.gov/abs/hep-ph/0401214}{{\tt hep-ph/0401214}}].

\bibitem{Kilian:2003xt}
W.~Kilian and J.~Reuter, {\it {The low-energy structure of little Higgs
  models}},  {\em Phys. Rev.} {\bf D70} (2004) 015004,
  [\href{http://xxx.lanl.gov/abs/hep-ph/0311095}{{\tt hep-ph/0311095}}].

\bibitem{Buras:2004kq}
A.~J. Buras, A.~Poschenrieder, and S.~Uhlig, {\it {Particle antiparticle
  mixing, $\varepsilon_K$ and the unitarity triangle in the Littlest Higgs
  model}},  {\em Nucl. Phys.} {\bf B716} (2005) 173--198,
  [\href{http://xxx.lanl.gov/abs/hep-ph/0410309}{{\tt hep-ph/0410309}}].

\bibitem{Choudhury:2004bh}
S.~R. Choudhury, N.~Gaur, A.~Goyal, and N.~Mahajan, {\it {$B_d - \bar B_d$ mass
  difference in little Higgs model}},  {\em Phys. Lett.} {\bf B601} (2004)
  164--170, [\href{http://xxx.lanl.gov/abs/hep-ph/0407050}{{\tt
  hep-ph/0407050}}].

\bibitem{Buras:2005iv}
A.~J. Buras, A.~Poschenrieder, and S.~Uhlig, {\it {Non-decoupling effects of
  the heavy $T$ in the $B_{d,s}^0 -\bar B_{d,s}^0$ mixing and rare $K$ and $B$
  decays}},  \href{http://xxx.lanl.gov/abs/hep-ph/0501230}{{\tt
  hep-ph/0501230}}.

\bibitem{Huo:2003vd}
W.-j. Huo and S.-h. Zhu, {\it {$b \to s \gamma$ in littlest Higgs model}},
  {\em Phys. Rev.} {\bf D68} (2003) 097301,
  [\href{http://xxx.lanl.gov/abs/hep-ph/0306029}{{\tt hep-ph/0306029}}].

\bibitem{Buras:2006wk}
A.~J. Buras, A.~Poschenrieder, S.~Uhlig, and W.~A. Bardeen, {\it {Rare $K$ and
  $B$ decays in the Littlest Higgs model without T- parity}},  {\em JHEP} {\bf
  11} (2006) 062, [\href{http://xxx.lanl.gov/abs/hep-ph/0607189}{{\tt
  hep-ph/0607189}}].

\bibitem{Cheng:2003ju}
H.-C. Cheng and I.~Low, {\it {TeV symmetry and the little hierarchy problem}},
  {\em JHEP} {\bf 09} (2003) 051,
  [\href{http://xxx.lanl.gov/abs/hep-ph/0308199}{{\tt hep-ph/0308199}}].

\bibitem{Cheng:2004yc}
H.-C. Cheng and I.~Low, {\it {Little hierarchy, little Higgses, and a little
  symmetry}},  {\em JHEP} {\bf 08} (2004) 061,
  [\href{http://xxx.lanl.gov/abs/hep-ph/0405243}{{\tt hep-ph/0405243}}].

\bibitem{Low:2004xc}
I.~Low, {\it {T parity and the littlest Higgs}},  {\em JHEP} {\bf 10} (2004)
  067, [\href{http://xxx.lanl.gov/abs/hep-ph/0409025}{{\tt hep-ph/0409025}}].

\bibitem{Hubisz:2005tx}
J.~Hubisz, P.~Meade, A.~Noble, and M.~Perelstein, {\it {Electroweak precision
  constraints on the littlest Higgs model with T parity}},  {\em JHEP} {\bf 01}
  (2006) 135, [\href{http://xxx.lanl.gov/abs/hep-ph/0506042}{{\tt
  hep-ph/0506042}}].

\bibitem{Asano:2006nr}
M.~Asano, S.~Matsumoto, N.~Okada, and Y.~Okada, {\it {Cosmic positron signature
  from dark matter in the littlest Higgs model with T-parity}},  {\em Phys.
  Rev.} {\bf D75} (2007) 063506,
  [\href{http://xxx.lanl.gov/abs/hep-ph/0602157}{{\tt hep-ph/0602157}}].

\bibitem{Cabibbo:1963yz}
N.~Cabibbo, {\it {Unitary Symmetry and Leptonic Decays}},  {\em Phys. Rev.
  Lett.} {\bf 10} (1963) 531--532.

\bibitem{Kobayashi:1973fv}
M.~Kobayashi and T.~Maskawa, {\it {CP Violation in the Renormalizable Theory of
  Weak Interaction}},  {\em Prog. Theor. Phys.} {\bf 49} (1973) 652--657.

\bibitem{Blanke:2006xr}
M.~Blanke {\em et.~al.}, {\it {Another look at the flavour structure of the
  Littlest Higgs model with T-parity}},  {\em Phys. Lett.} {\bf B646} (2007)
  253--257, [\href{http://xxx.lanl.gov/abs/hep-ph/0609284}{{\tt
  hep-ph/0609284}}].

\bibitem{Buras:2000dm}
A.~J. Buras, P.~Gambino, M.~Gorbahn, S.~Jager, and L.~Silvestrini, {\it
  Universal unitarity triangle and physics beyond the standard model},  {\em
  Phys. Lett.} {\bf B500} (2001) 161--167,
  [\href{http://xxx.lanl.gov/abs/hep-ph/0007085}{{\tt hep-ph/0007085}}].

\bibitem{Buras:2003jf}
A.~J. Buras, {\it Minimal flavor violation},  {\em Acta Phys. Polon.} {\bf B34}
  (2003) 5615--5668, [\href{http://xxx.lanl.gov/abs/hep-ph/0310208}{{\tt
  hep-ph/0310208}}].

\bibitem{D'Ambrosio:2002ex}
G.~D'Ambrosio, G.~F. Giudice, G.~Isidori, and A.~Strumia, {\it {Minimal flavour
  violation: An effective field theory approach}},  {\em Nucl. Phys.} {\bf
  B645} (2002) 155--187, [\href{http://xxx.lanl.gov/abs/hep-ph/0207036}{{\tt
  hep-ph/0207036}}].

\bibitem{Chivukula:1987py}
R.~S. Chivukula and H.~Georgi, {\it Composite Technicolor Standard Model},
  {\em Phys. Lett.} {\bf B188} (1987) 99.

\bibitem{Hall:1990ac}
L.~J. Hall and L.~Randall, {\it Weak scale effective supersymmetry},  {\em
  Phys. Rev. Lett.} {\bf 65} (1990) 2939--2942.

\bibitem{Hubisz:2005bd}
J.~Hubisz, S.~J. Lee, and G.~Paz, {\it {The flavor of a little Higgs with
  T-parity}},  {\em JHEP} {\bf 06} (2006) 041,
  [\href{http://xxx.lanl.gov/abs/hep-ph/0512169}{{\tt hep-ph/0512169}}].

\bibitem{Blanke:2006sb}
M.~Blanke {\em et.~al.}, {\it {Particle antiparticle mixing, $\varepsilon_K$,
  $\Delta\Gamma_q$, $A_\text{SL}^q$, $A_\text{CP}(B_d \to \psi K_S)$,
  $A_\text{CP}(B_s \to \psi \phi)$ and $B \to X_{s,d} \gamma$ in the Littlest
  Higgs model with T- parity}},  {\em JHEP} {\bf 12} (2006) 003,
  [\href{http://xxx.lanl.gov/abs/hep-ph/0605214}{{\tt hep-ph/0605214}}].

\bibitem{HongSheng:2007ve}
H.~Hong-Sheng, {\it {Flavor-changing top quark rare decays in the littlest
  Higgs model with T-parity}},  {\em Phys. Rev.} {\bf D75} (2007) 094010,
  [\href{http://xxx.lanl.gov/abs/hep-ph/0703067}{{\tt hep-ph/0703067}}].


\bibitem{Choudhury:2006sq}
S.~R. Choudhury, A.~S. Cornell, A.~Deandrea, N.~Gaur, and A.~Goyal, {\it
  {Lepton flavour violation in the little Higgs model}},  {\em Phys. Rev.} {\bf
  D75} (2007) 055011, [\href{http://xxx.lanl.gov/abs/hep-ph/0612327}{{\tt
  hep-ph/0612327}}].

\bibitem{Randall:1999ee}
L.~Randall and R.~Sundrum, {\it A large mass hierarchy from a small extra
  dimension},  {\em Phys. Rev. Lett.} {\bf 83} (1999) 3370--3373,
  [\href{http://xxx.lanl.gov/abs/hep-ph/9905221}{{\tt hep-ph/9905221}}].

\bibitem{Agashe:2004cp}
K.~Agashe, G.~Perez, and A.~Soni, {\it Flavor structure of warped extra
  dimension models},  {\em Phys. Rev.} {\bf D71} (2005) 016002,
  [\href{http://xxx.lanl.gov/abs/hep-ph/0408134}{{\tt hep-ph/0408134}}].

\bibitem{Gamiz:2006sq}
{\bf HPQCD} Collaboration, E.~Gamiz {\em et.~al.}, {\it {Unquenched
  determination of the kaon parameter $B_K$ from improved staggered fermions}},
   {\em Phys. Rev.} {\bf D73} (2006) 114502,
  [\href{http://xxx.lanl.gov/abs/hep-lat/0603023}{{\tt hep-lat/0603023}}].

\bibitem{Antonio:2007pb}
{\bf RBC} Collaboration, D.~J. Antonio {\em et.~al.}, {\it {Neutral kaon mixing
  from 2+1 flavor domain wall QCD}},  {\em Phys. Rev. Lett.} {\bf 100} (2008)
  032001, [\href{http://xxx.lanl.gov/abs/hep-ph/0702042}{{\tt
  hep-ph/0702042}}].

\bibitem{Aoki:2008ss}
{\bf JLQCD} Collaboration, S.~Aoki {\em et.~al.}, {\it {$B_K$ with two flavors
  of dynamical overlap fermions}},
  \href{http://xxx.lanl.gov/abs/0801.4186}{{\tt arXiv:0801.4186}}.

\bibitem{UTfitNEW}
{\bf UTfit} Collaboration. {\it Private communication}.

\bibitem{Lunghi:2008aa}
E.~Lunghi and A.~Soni, {\it {Possible Indications of New Physics in
  $B_d$-mixing and in $\sin(2 \beta)$ Determinations}},
  \href{http://xxx.lanl.gov/abs/0803.4340}{{\tt arXiv:0803.4340}}.

\bibitem{Buras:2008nn}
A.~J. Buras and D.~Guadagnoli, {\it {Correlations among new CP violating
  effects in $\Delta F = 2$ observables}},
  \href{http://xxx.lanl.gov/abs/0805.3887}{{\tt arXiv:0805.3887}}.

\bibitem{Aaltonen:2007he}
{\bf CDF} Collaboration, T.~Aaltonen {\em et.~al.}, {\it {First Flavor-Tagged
  Determination of Bounds on Mixing- Induced CP Violation in $B^0_s \to J/\psi
  \phi$ Decays}},  \href{http://xxx.lanl.gov/abs/0712.2397}{{\tt
  arXiv:0712.2397}}.

\bibitem{:2008fj}
{\bf D0} Collaboration, V.~M. Abazov {\em et.~al.}, {\it {Measurement of
  $\boldmath {B_s^0}$ mixing parameters from the flavor-tagged decay $B^0_s \to
  J/\psi \phi$}},  \href{http://xxx.lanl.gov/abs/0802.2255}{{\tt
  arXiv:0802.2255}}.

\bibitem{Bona:2008jn}
{\bf UTfit} Collaboration, M.~Bona {\em et.~al.}, {\it {First Evidence of New
  Physics in $b \leftrightarrow s$ Transitions}},
  \href{http://xxx.lanl.gov/abs/0803.0659}{{\tt arXiv:0803.0659}}.

\bibitem{Buchalla:1994tr}
G.~Buchalla and A.~J. Buras, {\it $\sin2\beta$ from $K \to \pi \nu\bar\nu$},
  {\em Phys. Lett.} {\bf B333} (1994) 221--227,
  [\href{http://xxx.lanl.gov/abs/hep-ph/9405259}{{\tt hep-ph/9405259}}].

\bibitem{Buras:2003td}
A.~J. Buras, {\it {Relations between $\Delta M_{s,d}$ and $B_{s,d} \to \mu^+
  \mu^-$ in models with minimal flavour violation}},  {\em Phys. Lett.} {\bf
  B566} (2003) 115--119, [\href{http://xxx.lanl.gov/abs/hep-ph/0303060}{{\tt
  hep-ph/0303060}}].

\bibitem{Hubisz:2004ft}
J.~Hubisz and P.~Meade, {\it {Phenomenology of the littlest Higgs with
  T-parity}},  {\em Phys. Rev.} {\bf D71} (2005) 035016,
  [\href{http://xxx.lanl.gov/abs/hep-ph/0411264}{{\tt hep-ph/0411264}}].

\bibitem{Flavianet}
{\bf Flavianet} Collaboration, {Kaon Working Group}. {{\tt
  http://www.lnf.infn.it/wg/vus/}}.

\bibitem{Yao:2006px}
{\bf Particle Data Group} Collaboration, W.~M. Yao {\em et.~al.}, {\it {Review
  of particle physics}},  {\em J. Phys.} {\bf G33} (2006) 1--1232. {Updates
  available on \texttt{http://pdg.lbl.gov}.}

\bibitem{Herrlich:1993yv}
S.~Herrlich and U.~Nierste, {\it {Enhancement of the $K_L - K_S$ mass
  difference by short distance QCD corrections beyond leading logarithms}},
  {\em Nucl. Phys.} {\bf B419} (1994) 292--322,
  [\href{http://xxx.lanl.gov/abs/hep-ph/9310311}{{\tt hep-ph/9310311}}].

\bibitem{Barberio:2007cr}
{\bf Heavy Flavor Averaging Group (HFAG)} Collaboration, E.~Barberio {\em
  et.~al.}, {\it {Averages of $b$-hadron properties at the end of 2006}},
  \href{http://xxx.lanl.gov/abs/0704.3575}{{\tt arXiv:0704.3575}}. {Updates
  available on \texttt{http://www.slac.stanford.edu/xorg/hfag}.}

\bibitem{Herrlich:1995hh}
S.~Herrlich and U.~Nierste, {\it {Indirect CP violation in the neutral kaon
  system beyond leading logarithms}},  {\em Phys. Rev.} {\bf D52} (1995)
  6505--6518, [\href{http://xxx.lanl.gov/abs/hep-ph/9507262}{{\tt
  hep-ph/9507262}}].

\bibitem{Herrlich:1996vf}
S.~Herrlich and U.~Nierste, {\it {The Complete $|\Delta S|=2$ Hamiltonian in
  the Next-To-Leading Order}},  {\em Nucl. Phys.} {\bf B476} (1996) 27--88,
  [\href{http://xxx.lanl.gov/abs/hep-ph/9604330}{{\tt hep-ph/9604330}}].

\bibitem{Buras:1990fn}
A.~J. Buras, M.~Jamin, and P.~H. Weisz, {\it {Leading and next-to-leading QCD
  corrections to $\epsilon$ parameter and $B^0 -\bar B^0$ mixing in the
  presence of a heavy top quark}},  {\em Nucl. Phys.} {\bf B347} (1990)
  491--536.

\bibitem{Urban:1997gw}
J.~Urban, F.~Krauss, U.~Jentschura, and G.~Soff, {\it {Next-to-leading order
  QCD corrections for the $B^0 - \bar B^0$ mixing with an extended Higgs
  sector}},  {\em Nucl. Phys.} {\bf B523} (1998) 40--58,
  [\href{http://xxx.lanl.gov/abs/hep-ph/9710245}{{\tt hep-ph/9710245}}].

\bibitem{Bona:2007mn}
M.~Bona {\em et.~al.}, {\it {Status of the Cabibbo-Kobayashi-Maskawa matrix and
  unitarity triangle fits}},  {\em AIP Conf. Proc.} {\bf 881} (2007) 210--219.

\bibitem{Buras:2005hz}
A.~J. Buras, R.~Fleischer, S.~Recksiegel, and F.~Schwab, {\it {Electroweak
  penguin hunting through $B \to \pi \pi, \pi K$ and rare $K$ and $B$ decays}},
   {\em PoS} {\bf HEP2005} (2006) 193,
  [\href{http://xxx.lanl.gov/abs/hep-ph/0512059}{{\tt hep-ph/0512059}}].

\bibitem{Blanke:2006ig}
M.~Blanke, A.~J. Buras, D.~Guadagnoli, and C.~Tarantino, {\it {Minimal Flavour
  Violation Waiting for Precise Measurements of $\Delta M_s$, $S_{\psi \phi}$,
  $A^s_\text{SL}$, $|V_{ub}|$, $\gamma$ and $B^0_{s,d} \to \mu^+ \mu^-$}},
  {\em JHEP} {\bf 10} (2006) 003,
  [\href{http://xxx.lanl.gov/abs/hep-ph/0604057}{{\tt hep-ph/0604057}}].

\bibitem{Adler:2001xv}
{\bf E787} Collaboration, S.~S. Adler {\em et.~al.}, {\it {Further evidence for
  the decay $K^+ \to\pi^+ \nu\bar\nu$}},  {\em Phys. Rev. Lett.} {\bf 88}
  (2002) 041803, [\href{http://xxx.lanl.gov/abs/hep-ex/0111091}{{\tt
  hep-ex/0111091}}].

\bibitem{Anisimovsky:2004hr}
{\bf E949} Collaboration, V.~V. Anisimovsky {\em et.~al.}, {\it {Further study
  of the decay $K^+\to\pi^+\nu\bar\nu$}},  {\em Phys. Rev. Lett.} {\bf 93}
  (2004) 031801, [\href{http://xxx.lanl.gov/abs/hep-ex/0403036}{{\tt
  hep-ex/0403036}}].

\bibitem{Grossman:1997sk}
Y.~Grossman and Y.~Nir, {\it {$K_L\to\pi^0\nu\bar\nu$ beyond the standard
  model}},  {\em Phys. Lett.} {\bf B398} (1997) 163--168,
  [\href{http://xxx.lanl.gov/abs/hep-ph/9701313}{{\tt hep-ph/9701313}}].

\bibitem{Buras:2001af}
A.~J. Buras and R.~Fleischer, {\it {Bounds on the unitarity triangle,
  $\sin2\beta$ and $K \to\pi \nu\bar\nu$ decays in models with minimal flavor
  violation}},  {\em Phys. Rev.} {\bf D64} (2001) 115010,
  [\href{http://xxx.lanl.gov/abs/hep-ph/0104238}{{\tt hep-ph/0104238}}].

\end{thebibliography}\endgroup
%\bibliographystyle{JHEP}

\end{document}